\documentstyle[prl,aps,preprint,tighten,floats,aps,epsf,psfig]{revtex}

\def\simgt{\stackrel{>}{{}_\sim}}
\def\be{\begin{equation}}
\def\ee{\end{equation}}
\def\bear{\begin{eqnarray}}
\def\eear{\end{eqnarray}}

\def\VEV#1{{\langle #1 \rangle} }

\begin{document}


\draft
\preprint{\vbox{\baselineskip=12pt
\rightline{CERN-TH/2002-026}
\vskip0.2truecm
\rightline{MCTP-02-03}
\vskip0.2truecm
\rightline{SHEP 02-03}
%
\vskip0.2truecm
\rightline{hep-ph/0202100}}}

\title{Supersymmetric Pati-Salam Models from Intersecting D-Branes}
\author{L. L.
Everett${}^{1,2}$, G. L. Kane${}^{2}$, S. F. King${}^{3}$, S.
Rigolin${}^{1,2}$, and
Lian-Tao Wang${}^{2}$}
\address{1. CERN, TH Division, CH-1211 Geneva 23, Switzerland} 
\address{2. Michigan Center for Theoretical Physics, Ann Arbor, Michigan,
48109, USA}
\address{3. Department of Physics and
Astronomy, University of Southampton\\ Southampton, S017 1BJ, UK} 
\maketitle 

\begin{abstract} 
We explore supersymmetric Type I string-motivated three-family scenarios
in which the Standard Model is embedded within two sets of intersecting D
branes with $U(1)$-extended Pati-Salam gauge groups. We study a model 
inspired by the Shiu-Tye Type IIB orientifold, in which a three-family
scenario is obtained by assuming that the gauge symmetry breaking takes
place in two stages; the Pati-Salam group arises from diagonal breaking of
the $U(N)$ gauge groups, which is then broken to the SM gauge group.  We
investigate the diagonal breaking scenario in detail and find that
generically there are difficulties involved in decoupling the exotic Higgs
remnants.  On the phenomenological side, proper low 
energy gauge coupling predictions 
effectively lead to a ``single brane'' scenario for
$M_{String}\sim M_U\simeq 10^{16}$ GeV. The soft parameters in this limit 
are constrained by a well-known sum rule, leading to a distinctive
phenomenological pattern for the low energy mass spectrum.
\end{abstract}

\newpage

\section{Introduction}
One of the most important unsolved questions for superstring theory is to
determine the way(s) in which it can be connected to our observable, low
energy world.  The main obstacles have been well-known since the
development of perturbative string theory: (i) the tremendous degeneracy
of string vacua and (ii) how to break supersymmetry (SUSY) and stabilize
moduli while resolving the hierarchy problem.  
Recent developments in string theory have not suggested definitive
resolutions to either issue, although they have provided new settings in
which to investigate such questions.

Given the absence of a dynamical principle for selecting the correct
string vacuum, one may argue that the
question of how the Standard Model (SM) or some plausible extension such
as the minimal supersymmetric standard model (MSSM) emerges should be put
on hold until string theory dynamics are further understood.  We
instead suggest that while understanding the dynamics of vacuum selection 
is indeed the ultimate goal, the answer is not likely to be obtainable 
solely by studying the formal structure of string theory.
It is more likely that progress will follow the historical path, in which 
low energy data provides invaluable guidance in uncovering the nature of
the high energy theory.

In our view it is thus important to consider the phenomenological
implications of four-dimensional compactifications of string theory.  
The Type I string framework has particular
interest in that it can provide explicit realizations of the
braneworld, both with low fundamental scale of $O($TeV) 
and higher fundamental scales ranging from intermediate 
$O(10^{11}$ GeV) to GUT values. Many string constructions have been 
explored, both  with supersymmetry   
\cite{sagnotti,berkooz,aldazabal1,kakushadze,shiu,cvetic,rabadan1,csu} and 
without \cite{aldazabal2,rabadan2}.
The nonsupersymmetric models have quite 
realistic features, but their stability is in doubt.
The supersymmetric constructions have stable D-brane configurations but 
generically are not phenomenologically 
appealing. There has recently been progress in 
model building which has led to more realistic models \cite{csu}.

In this paper, we study supersymmetric Type I string-motivated models in 
which the SM gauge group is split between 
different D-brane sectors.  
The extension we consider is inspired by an explicit string construction, 
the Shiu-Tye model \cite{shiu}. This model 
has two sets of intersecting D-branes 
which each have a copy of a $U(1)$-extended Pati-Salam 
\cite{patisalam} gauge group \( 
U(4)_c\times U(2)_{L}\times
U(2)_{R} \). The right-handed neutrinos are present
as open string states in a natural way (in the 
same multiplet as the quark and charged lepton singlets) as expected in
Pati-Salam models.
As in the Shiu-Tye model, the SM gauge group is embedded 
in a nontrivial
way: the Pati-Salam gauge groups which break to the SM are obtained from a
diagonal breaking of the gauge groups of the two D-brane sectors. The
origin of the three SM families is connected to the gauge embedding; the 
third family and the electroweak Higgs doublets are open string states
from one set of D-branes, while the first and second generations
are open string states with endpoints on each of the D-brane sectors.  

Such scenarios are interesting because the Pati-Salam framework (which 
has been extensively studied in the literature) provides a setting 
in which to explore many phenomenological issues.
Nonsupersymmetric Pati-Salam models motivated from D-brane setups have 
also been recently investigated \cite{leontaris}.  We analyze 
supersymmetric Pati-Salam models motivated from intersecting D-brane 
scenarios. In doing so, we 
assume for definiteness that the string scale 
is approximately equal to the unification scale $M_{String}\sim 
M_{U}\simeq
10^{16}$ GeV ({\it i.e.}, all compactification radii are small), which 
is also the Pati-Salam symmetry breaking scale.
This high string scale is chosen to make contact 
with the
successes of traditional SUSY GUTs; however, it may not 
be preferred within generic Type I models.\footnote{The 
presence of antibranes (hence nonsupersymmetric sectors) may be required 
to satisfy tadpole/anomaly 
cancellation constraints \cite{burgess,ibanezquevedo,abelibanez}. The 
string scale should then be
lowered to intermediate values if the SUSY-breaking sector is hidden from 
the observable brane(s) and lowered to $O({\rm TeV})$ if the
SUSY-breaking sector has direct couplings to the observable sector.} In
this case, obtaining proper low energy values of the gauge couplings
effectively leads to a ``single brane'' model. In this limit the soft
terms are constrained by a sum rule well known within perturbative 
heterotic orbifold models. We explore this scenario within a concrete 
model
(``4224'')
in which the \(U(2)_{L,R} \) arise from one set of branes, while $U(4)_c$
is obtained from diagonal breaking of the $U(4)$ groups from
each D-brane sector. While the model has interesting
features, there are phenomenological difficulties which will 
be explained in detail. 

\section{Theoretical Framework}

The superstring-motivated models we consider are based on $d=4$, $N=1$
orientifold compactifications of Type II superstring theory
\cite{sagnotti,berkooz,aldazabal1,kakushadze,shiu,cvetic,rabadan1,csu}.  
In \cite{ibanez} the general features of this class of models as well as
details of the effective action at string tree level of the massless
states of the open string sectors are presented.\footnote{Note this
effective action was derived using T duality arguments \cite{ibanez} 
valid for 
models with all D-branes are
located at a single fixed point (not models for which subsets of
branes are located at different fixed points or with nontrivial
backgrounds).} 
We consider scenarios in which
the non-Abelian gauge groups of the theory are due to two intersecting
(orthogonal) stacks of D five-branes (labeled $5_1$ and $5_2$) located at
a single orbifold fixed point.

The chiral matter fields (from which the MSSM matter fields will be
obtained) and gauge bosons arise from the open string sectors, while the
dilaton, moduli, and graviton fields are in the closed string 
sector.\footnote{Here we shall ignore the presence of other closed 
string states, the twisted moduli.  Such states play interesting roles 
in the theoretical structure of the model and can lead to 
distinctive phenomenological signatures (see e.g.    
\cite{benakli,abelibanez,krtwist}).}
The open string states are labeled according to the location of the string
endpoints on the different D-branes of the theory. These states are
classified into two categories; the notation of \cite{ibanez} is used 
to label them. The first includes open strings which start and end on the
same set of branes \(C_{j}^{5_{1,2}} \) (where $j$ labels the three
complex compact dimensions). These fields are in fundamental or
antisymmetric tensor representations of the non-Abelian gauge group of the 
corresponding set of
branes.  The second category includes open strings which start and end on
different sets of branes \( C^{5_{1}5_{2}} \). These fields are
bifundamentals with respect to the two non-Abelian gauge groups.  Note 
that within
this class of models, non-Abelian gauge singlet fields with Abelian
charges are generically absent (for stacks of D-branes with $U(N)$ gauge 
groups).

An example of a three family orientifold model with the SM gauge group is 
the Shiu-Tye model \cite{shiu}, a $Z_6$ model with
background NS-NS field.
In this model, the $5_1$ sector has gauge group 
$U(4)^{(1)}\times U(2)^{(1)} \times U(2)^{(1)}$ and the $5_2$ sector has 
$U(4)^{(2)}\times U(2)^{(2)} \times U(2)^{(2)}$
(the existence of identical
gauge groups on each set of branes is guaranteed by T duality). 
The gauge group on each brane is a U(1)-extended version 
of the Pati-Salam model ($SU(4)_c\times SU(2)_L\times SU(2)_R$). 
Recall in the Pati-Salam model, the MSSM states lie in the multiplets 
\(F_{a}=(4,2,1) \) and \(
\bar{F}_{a}=(\bar{4},1,2) \),
while the Higgs doublets \(
H_{u,d} \) lie in the bi-doublet \( h=(1,2,2) \). The Higgs fields which
break the Pati-Salam gauge group to $SU(3)\times SU(2)_L\times U(1)_Y$  
(in which the hypercharge originates from the breaking of the
non-Abelian gauge structure) are a pair of vectorlike states \(
H=(4,1,2) \) and \( \bar{H}=(\bar{4},1,2) \). 
The right-handed neutrinos have a natural origin
in the theory as components of the $\bar{F}_a$ multiplets.

The three families arise in the Shiu-Tye model in an interesting way. The
massless spectrum of this model contains one chiral family associated
with each of the two D-brane sectors and two families arising from the
intersection of the D-brane sectors.  Three family scenarios thus
require that the SM gauge group be embedded in a nontrivial way within
the two D-brane sectors. The desired embedding is that $U(2)_L$
and $U(2)_R$ each arise from the diagonal breaking of one set of
$U(2)^{(1)}$ and $U(2)^{(2)}$ from the $5_1$ and $5_2$ sectors. The
symmetry breaking is triggered by diagonal vacuum expectation values
(VEV's) of bifundamental fields charged with respect to the $U(2)$ groups
of each sector. $SU(3)_c$ is obtained from (say) the breaking of
$U(4)^{(1)}$, such that the third family arises from the $5_1$ sector.

In the Shiu-Tye model, the doubling of the intersection states is
connected with the nonvanishing quantized background NS-NS B
field. The presence of the background B field modifies the vertex
operators of the intersection states, allowing for a nontrivial
multiplicity of states.  Such states are not identical from the worldsheet
point of view and the selection rules which govern their couplings
should be different from those presented for orientifolds without
nontrivial background field VEV's in \cite{berkooz,ibanez}.  Given the
class of available results we shall use the effective action which treats
these two families identically, but caution the reader that this symmetry 
is not likely to hold to all orders in the effective action.

\section{4224 Model}

Motivated by the Shiu-Tye model, we consider a different embedding of 
the SM gauge group from intersecting five-branes with Pati-Salam gauge 
groups. The gauge group of the $5_{1}$ sector is
$U(4)^{(1)}\times U(2)_{L}\times U(2)_{R} $ and the gauge group of the 
$5_{2}$
sector is \( U(4)^{(2)} \) (e.g., we assume the $U(2)$ gauge groups of the 
$5_2$ 
sector are broken).  In this ``4224'' model, the third family arises from 
the $5_{1}$ sector, while the first and  second families are intersection 
states. The particle content is given in Table I.

\begin{table}[tbp]
\begin{center}
\begin{tabular}{||c||c|c|c|c||c|c|c|c||c||}
\hline
 & &  & & & & & & &\\
& $SU(4)^{(1)}$& $SU(2)_L$ 
&$SU(2)_R$&$SU(4)^{(2)}$&$Q^{(1)}_4$&$Q_{2L}$&$Q_{2R}$&$Q^{(2)}_4$ 
&  \\
\hline
$h$ &1&2&2&1  & 0&$1$&$-1$&0& $C^{5_1}_1$\\
$F_3$&4&2&1&1 & $1$&$-1$&0&0& $C^{5_1}_2$\\
$\bar{F}_3$ &$\bar{4}$&1&2&1&$-1$&0&1&0& $C^{5_1}_3$ \\
$F_{1,2}$ &1&2&1&4& 0&$-1$&0&1&$C^{5_15_2}$\\
$\bar{F}_{1,2}$&1&1&2&$\bar{4}$ & 0&0&1&$-1$& $C^{5_15_2}$\\
$H$ & 4&1&2&1& $1$&0&$-1$&0& $C^{5_1}_1$\\
$\bar{H}$&$\bar{4}$&1&2&1 & $-1$&0&1&0&$C^{5_1}_2$\\
$\varphi_1$&4&1&1&$\bar{4}$ &1&0&0&$-1$ & $C^{5_15_2}$\\
$\varphi_2$&$\bar{4}$&1&1&4 &$-1$&0&0&1 & $C^{5_15_2}$\\
$D^{(+)}_6$ & 6&1&1&1& $2$&0&$0$&0& $C^{5_1}_1$\\
$D^{(-)}_6$&$6$&1&1&1 & $-2$&0&0&0&$C^{5_1}_2$\\
\hline 
\end{tabular}
\end{center}
\caption{The particle content of the 4224 model.}
\end{table} 

Let us first discuss the additional 
$U(1)$ factors. A close inspection of Table I shows that all but one 
linear combination of 
the $U(1)$'s ($U(1)'=Q^{1}_4+Q^{2}_4+Q_{2L}+Q_{2R}$) are anomalous (see 
also \cite{leontaris}) and  
will be broken at the string scale by the 4D Green Schwarz 
mechanism 
of Type I \cite{anom}.\footnote{In 
contrast to the 
perturbative heterotic models which have at most one anomalous $U(1)$ 
since it is the dilaton which participates in the anomaly cancellation, 
within Type I models there can be multiple anomalous $U(1)$'s which are 
cancelled by shifts in the twisted moduli fields.}  
In the 4224 model all of the SM fields 
listed in Table I as well as the Higgs fields $\{H, 
\bar{H},\varphi_{1,2}\}$ are  neutral under the persisting nonanomalous 
$U(1)'$.
The only fields charged under this $U(1)$ are the $D_6$ fields 
needed to decouple the Higgs triplets after the Pati-Salam 
gauge symmetry breaking. 
As these 
fields must acquire superheavy masses to avoid fast proton decay,   
this $U(1)'$ (although it remains unbroken unless additional 
fields are introduced into the model to break it) does not have an 
important impact on the low energy phenomenology. 
However, the presence of the $U(1)$ factors at the string scale requires 
that the effective action respect $U(N)$ symmetry rather than $SU(N)$; 
{\it i.e.} the  $U(1)$'s persist as global symmetries.  These effective 
global symmetries have important consequences. For example 
\cite{shiu,leontaris}, within the $U(1)$-extended Pati-Salam model one of 
the surviving global symmetries is $U(1)_B$, ensuring the 
stability of the proton even with low fundamental scales (assuming effects 
of the colored Higgses are suppressed, which will be shown to be a 
difficult task in the next subsection).

\subsection{Gauge Symmetry Breaking and Decoupling of Exotic Matter}

The gauge symmetry breaking pattern of this model follows the 
three-family approach of Shiu and Tye \cite{shiu}. The symmetry breaking 
takes place in two stages, which are assumed to occur at very similar 
scales $\sim M_U$. In the 
first stage, $U(4)^{(1)}\times U(4)^{(2)}$ are broken to the 
diagonal $U(4)$ subgroup (which is identified as $U(4)_c$) by diagonal 
VEV's of the bifundamental fields 
$\varphi_{1,2}$.  The resulting theory is an effective Pati-Salam model 
(with additional $U(1)$'s) which then breaks to the MSSM (and additional 
$U(1)$'s, which we shall ignore) through the 
Higgs pair of bifundamentals $H$, $\bar{H}$.
To preserve supersymmetry, the symmetry 
breaking should occur along flat directions; D flatness can be spoiled 
only by terms of the order of the soft supersymmetry breaking masses. 
Specifically, the symmetry breaking occurs as follows:
\begin{eqnarray}
U(4)^{(1)}\times U(4)^{(2)}\times U(2)_L\times U(2)_R 
\stackrel{\VEV{\varphi_{1,2}}}\longrightarrow  U(4)_c\times U(2)_L\times 
U(2)_R 
\nonumber\\ 
\stackrel{\VEV{{H,\bar{H}}}}\longrightarrow 
SU(3)_C\times 
SU(2)_L\times 
U(1)_Y\times U(1)^3.
\end{eqnarray}
It is instructive to consider this symmetry breaking pattern in further 
detail and investigate the decoupling of the exotic matter present in the 
Higgs representations.  

Let us first consider the diagonal breaking step, $U(4)\times 
U(4)\rightarrow U(4)_c$.  This is achieved when the bi-fundamental fields 
$\varphi_{1}$ and $\varphi_{2}$ acquire diagonal VEV's
\begin{equation}
\langle (\varphi_{1})_{\alpha a} \rangle=\delta_{\alpha a}v_1;\;\;\langle 
(\varphi_{2})_{\alpha a} \rangle=\delta_{\alpha a}v_2,
\end{equation}
in which $\alpha$ denotes the $U(4)^{(1)}$ gauge index and $a$ that of
$U(4)^{(2)}$. D flatness (particularly for the $U(1)$ generators 
of the $U(4)$ gauge groups) requires that
$v_1=v_2$ ($\sim M_U$).\footnote{The D
flatness conditions can be modified if the associated $U(1)$'s are
anomalous. In this case one can choose to resolve or ``blow up'' the
orbifold singularities by giving VEV's to the twisted moduli \cite{anom},
which induces nonzero Fayet-Iliopoulos D terms.  This highly  
model-dependent option is not considered further in this paper.}

Setting the effects of the soft supersymmetry breaking to zero for the
moment, one can easily understand the resulting pattern of Higgs masses.  
As the gauge symmetry is broken down to the diagonal subgroup but
supersymmetry is preserved, the supersymmetric completion of the Higgs
mechanism requires that the resulting fields arrange themselves into
supermultiplets.  There are 16 broken generators of the gauge group, while
the states $\varphi_{1,2}$ have a total of 32 (complex) degrees of
freedom. Of these 32 complex degrees of freedom in the Higgs sector, 16
real degrees of freedom are Goldstone bosons eaten to form the
longitudinal components of the massive gauge bosons. 16 additional real
degrees of freedom in the Higgs sector also obtain masses via D terms.
These fields are degenerate in mass with the gauge bosons and together
with the 16 massive fermions (with masses due to gaugino-Higgsino mixing)
form a massive vector supermultiplet. Therefore, 16 complex degrees of
freedom remain massless; one real degree of freedom corresponds to the
flat direction and there is also another Goldstone boson corresponding to
the global $U(1)$ symmetry which remains from the freedom to perform
global phase rotations on the fields $\varphi_{1,2}$ (the Goldstone boson
of the other global $U(1)$ is eaten to form one of the massive gauge
bosons).  The other massless fields organize themselves into an adjoint
representation of $SU(4)_c$, which consists of a color octet,
two color triplets, and one color singlet with respect to $SU(3)_c$.  If
such fields remain massless, the proton is not stable. This would
certainly rule out this model and suggests serious difficulties with this
intersecting brane approach.

In principle, a potential could be constructed 
utilizing higher-dimensional
operators which could presumably decouple these additional states (as well 
as stabilize the VEV's at the scale $M_U$). This is beyond the scope of 
this paper.
It is worth noting that if the gauge groups were 
$SU(N)$ rather 
than $U(N)$, D flatness only requires one Higgs field rather 
than the vectorlike pair (of course anomaly requirements
would require additional states in that case).  The number of
degrees of freedom in the Higgs sector is then halved and there is a 
desirable matching of the number of Higgs fields to the number of broken
generators.\footnote{This can also be
understood using the well-known correspondence between holomorphic gauge
invariant polynomials and D flat directions \cite{dflat}. In the $SU(N)$
case the invariant polynomial is given by $\varphi^N$, where $\varphi$ is
the bifundamental field with the diagonal VEV.  In the $U(N)$ case gauge
invariance forbids this term and requires the existence of an oppositely
charged scalar $\bar{\varphi}$, such that the holomorphic gauge invariant
polynomial is $\varphi \bar{\varphi}$.} 

In the second stage of symmetry breaking, the additional $U(1)$'s do not
complicate the D flatness analysis. However, complications arise for 
decoupling the additional color triplets
from $H,\bar{H}$ after the Pati-Salam symmetry breaking. The standard
mechanism \cite{antoniadis} of utilizing a single $SU(4)$ sixplet field 
$D_6$ which couples
to $H,\bar{H}$ via $W=D_6HH+D_6\bar{H}\bar{H}$ naively does not work
because the necessary operators (for one of the terms) are
forbidden by $U(4)$ gauge invariance (due to the $U(1)$ charges,
as the $D_6$ field can either have $U(1)$ charge $\pm 2$).  This
problem could be solved via the introduction of non-Abelian singlet
fields which are charged under the $U(1)$'s and acquire string-scale
VEV's. However, such fields are not expected in orientifold models when
the D-branes are all located at a single fixed point. One can
introduce both $D_6^{(+)}$ and $D^{(-)}_6$, in which case
gauge invariance allows $D^{(-)}_6 HH$ and $D^{(+)}_6 \bar{H}\bar{H}$
(although string selection rules still forbid these terms at trilinear
order). However, this will not eliminate massless color
triplets, as another color triplet pair has been introduced that 
remains massless barring the presence of additional couplings.

In the 4224 setup, the D-brane
assignments of $H$ and $\bar{H}$ also forbid a trilinear coupling with a 
$U(4)$ sixplet field $D_6$ from the $5_1$ brane sector (as the only
string-allowed coupling is of the form $C^{5_1}_1C^{5_1}_2C^{5_1}_3$).  
This problem could be solved if the fields which break the
Pati-Salam gauge groups are chosen to be a pair of $C^{5_15_2}$ states 
$H'=(1,1,\bar{2},4)$, $\bar{H}'=(1,1,2,\bar{4})$.
In this case the $D_6'=(1,1,1,6)$ field can be chosen as a $C^{5_2}_3$ 
field, which leads to a string-allowed trilinear coupling
$C^{5_2}_3C^{5_15_2}C^{5_15_2}$. However, in this case the baryon number
violating couplings $FFD_6$ and $\bar{F}\bar{F}D_6$ would be equally
allowed.\footnote{A global $R$ symmetry is typically utilized 
to forbid 
such terms \cite{kingshafi}.}  In principle, 
{\it both} sets of Higgs pairs $H$, $\bar{H}$, $H'$, $\bar{H}'$ could be
present. In this case, the number of
Higgs fields needed to break the same symmetry group has been 
doubled. This again leads to further massless states (additional color 
triplets of both the up and down type). 

To summarize, it is clear that this model suffers
from a major drawback in that it is difficult to decouple the exotic 
states.  As many of these states are charged under
$SU(3)_c$, they must acquire superheavy masses or they can lead to rapid
proton decay.  In this setup the additional $U(1)$'s are the source of the 
trouble.  This may have implications for such intersecting D-brane setups, 
since the $U(N)$ (rather than $SU(N)$) groups are generic to 
D-branes. The additional
$U(1)$'s make their presence felt in two ways.  First, D flatness
conditions necessarily lead to a doubling of the Higgs sector in the
diagonal breaking step.  Second, $U(1)$ gauge
invariance (and worldsheet selection rules) can forbid the necessary
couplings required to generate masses for the additional states.  The 
issues outlined above for the 4224 model will hold for generic 
supersymmetric setups with intersecting branes on orbifolds.  
Similar issues occur within the Shiu-Tye model, although there the exotic 
states  from the diagonal breaking are charged under 
$SU(2)_{L}$ and $U(1)_Y$ rather than $SU(3)_c$.

Having noted this generic (and problematic) feature, in the rest of the
paper we proceed to study the effective action and phenomenology of the 
MSSM sector of the 4224 model.  In other words, we assume that
mechanisms exist which decouple the unwanted exotic states at the string
scale. We then study the ``observable'' sector assuming that only the MSSM
states are present below the string scale.

\subsection{Effective Action of MSSM Fields}

We now discuss the form of the effective action
of the MSSM sector of the 4224 model, beginning with the 
gauge couplings.  
The symmetry breaking pattern leads to the 
following relations among the gauge couplings of the SM gauge groups in 
terms of the gauge couplings $g_{5_1}$ and $g_{5_2}$ associated with the 
$5_1$ and $5_2$ sectors (see also \cite{shiu}): 
\begin{eqnarray}
g_{3} & = & 
\frac{g_{5_{1}}g_{5_{2}}}{\sqrt{g^{2}_{5_{1}}+g^{2}_{5_{2}}}}=g_4\nonumber\\
g_{2} & = & g_{5_{1}} =g_{2R}\nonumber\\
g_{Y} & = & \frac{\sqrt{3}g_{3}g_{2}}{\sqrt{3g^{2}_{3}+2g^{2}_{2}}}.
\label{gaugecouplings1}
\end{eqnarray}
The gauge couplings are not unified when $g_{5_1}$ and $g_{5_2}$ are
unrelated, as discussed below.

We now turn to the structure of the superpotential Yukawa couplings.    
The gauge and D-brane assignments of the model are
designed to allow the $O(1)$ Yukawa coupling of the third family
\(F_{3}\bar{F}_{3}h \), as
$C^{5_1}_1C^{5_1}_2C^{5_1}_3$ is allowed by worldsheet symmetries. The 
model as expected
predicts third family Yukawa unification, such that large
of $\tan \beta$ is required to ensure the correct $m_t/m_b$ mass ratio.  
The first and second generations do not have large Yukawa
couplings at trilinear order as required by phenomenology.  The
selection rules presented in \cite{ibanez} dictate that 
$F_{1,2}\bar{F}_{1,2}h$ is forbidden 
with our choice of the Higgs D-brane assignment.  Gauge invariance also 
forbids terms of the form $F_{1,2}\bar{F}_3h$ and $F_{3}\bar{F}_{1,2}h$.

Therefore, the small Yukawa couplings of the first and second families as 
well as the Majorana mass terms for the
neutrinos must therefore be generated via higher-dimensional 
operators.  
In \cite{steve} these entries in the Yukawa matrix
were generated by operators
of the form $F_i\bar{F}_jh(H\bar{H})^n$. In the present model
there can be two pairs of heavy Higgs
fields $\{H,\bar{H},H',\bar{H}'$\}.
The operators $F_{1,2}\bar{F}_3h(H\bar{H}')$ are allowed by gauge 
invariance. If they are permitted by worldsheet symmetries, the 
Clebsch structure could allow
for a large 23 entry in the neutrino Yukawa matrix and simultaneously    
small 23 entries in the other Yukawa matrices,  allowing for a large 23
neutrino mixing angle \cite{MO2}.
The small 13 mixing angles would suggest that the 13 operators are
all suppressed by some additional symmetry which plays the   
role of the $U(1)$ family symmetry in \cite{MO2}.
It is also unclear whether unsuppressed operators such as
$F_{1,2}\bar{F}_{1,2}h(H\bar{H})$ are permitted.
Note that the addition of the two pairs of Higgs fields needed for 
interesting lepton textures lead to additional color triplet states which 
remain massless after the symmetry breaking unless
additional higher-dimensional operators are introduced to decouple them.

Thus, such intersecting D-brane models 
potentially have
a rich flavor structure, although the details of the higher-dimensional 
operators required remain speculative. To determine 
whether any flavor structure is viable will of course depend  in detail on 
these  
smaller Yukawa couplings (and the associated soft trilinear couplings).
Despite these limitations, it is worth
elaborating on certain aspects of the flavor structure which might be 
expected.
As stated, the 33 Yukawa entry gets a large
tree level contribution while the other entries are only generated by
higher order operators. Assuming that the first two generations are
governed by the same selection rules at leading order, the
Yukawas would take the following approximate form:
\begin{equation}
\label{yukorg}
Y_{ij}\simeq \left[ \begin{array}{ccc}
 a& a &b\\
 a& a &b\\
 c& c &1
\end{array}
\right] ,
\end{equation}
where $a$, $b$ and $c$ are generated by higher-dimensional operators.
Given the assumed (near) degeneracy of the first and second generation
gauge eigenstates, a hierarchy of fermion masses can naturally
be generated within this framework.

At this stage we can write down the soft parameters of the 4224
model, using the phenomenological approach of \cite{bim}.\footnote{Within 
this approach (with a high string scale $M_{String}\sim M_U \sim 10^{16}$ 
GeV), the presence of a hidden (which communicates with the 
observable sector only gravitationally) or sequestered sector 
is assumed in which the $N=1$ SUSY present at the string scale is broken 
by some unknown dynamical mechanism (i.e. gaugino condensation) within 
the effective field theory. We do not
attempt to model this sector in this paper. However, if
all the branes within the theory are assumed to sit on a single fixed
point, there is no truly sequestered sector.  It is not known whether 
potentially realistic orientifold models with $N=1$ SUSY can be 
constructed which have these relevant features (as discussed in 
\cite{ibanezquevedo}).} 
We use the notation
\begin{eqnarray}
F^S&=&\sqrt{3}(S+S^{*})m_{3/2} X_0 e^{i\alpha_S}\nonumber\\
F^i&=&\sqrt{3}(T_i+T^{*}_i)m_{3/2} X_i e^{i\alpha_i},
\end{eqnarray}
in which $m_{3/2}$ is the gravitino mass and $X_{0,i=1,2,3}$ measure the
relative contributions of $S$ and $T_i$ to the supersymmetry breaking
(the $X_i$ are related to the usual Goldstino angle
parameterization \cite{bim,ibanez} via $X_0\equiv \sin\theta$, $X_i
\equiv \cos\theta \Theta_i$). The $X_i$ satisfy the relation $\sum_i
X_i^2=1$. In general, the $F$- component VEV's can have
arbitrary complex phases $\alpha_S$, $\alpha_i$.  The gaugino masses of 
the Pati-Salam gauge groups are given by:
\begin{eqnarray}
m^{(1)}_{4}&=&\sqrt{3}m_{3/2}X_{1}e^{-i\alpha 
_{1}}=m_{2L}=m_{2R}\nonumber\\
m^{(2)}_{4}&=&\sqrt{3}m_{3/2}X_{2}e^{-i\alpha _{2}},
\end{eqnarray}
such that the MSSM gaugino masses are 
\begin{eqnarray}
\label{m3}
m_{3} & = & \frac{\sqrt{3}m_{3/2}}{(T_{1}+
T^{*}_{1})+(T_{2}+T^{*}_{2})}[(T_{1}+ T^{*}_{1})X_{1}e^{-i\alpha
_{1}}+(T_{2}+T^{*}_{2})X_{2}e^{-i\alpha _{2}}]\nonumber\\   
m_{2} & = & \sqrt{3}m_{3/2}X_{1}e^{-i\alpha _{1}}\nonumber\\
m_{1} & = & \frac{\sqrt{3}m_{3/2}}{\frac{5}{3}(T_{1}+T^{*}_{1})+
\frac{2}{3}(T_{2}+T^{*}_{2})}[\frac{5}{3}(T_{1}+
T^{*}_{1})X_{1}e^{-i\alpha _{1}}+\frac{2}{3}(T_{2}+
T^{*}_{2})X_{2}e^{-i\alpha _{2}}]. 
\end{eqnarray}
The soft mass-squared parameters are given by 
\begin{eqnarray}
m_{h}^{2} & =& m^2_{H_u} = m^2_{H_d}= m_{3/2}^{2}(1-3X^{2}_{0})\\
m_{F_{3}}^{2} & = & m^2_{Q_3}=m^2_{L_3}=m_{3/2}^{2}(1-3X^{2}_{3})\\
m_{\bar{F}_{3}}^{2} & = &
m^2_{U_3}=m^2_{D_3}=m^2_{E_3}=m_{3/2}^{2}(1-3X^{2}_{2})\\
m_{F_{1,2}}^{2}&=&m_{Q_{1,2}}^{2}=m^2_{L_{1,2}}  =  m_{3/2}^{2}(1-
\frac{3}{2}(X^{2}_{0}+X^{2}_{3}))\\
m_{\bar{F}_{1,2}}^{2}&=&m^2_{U_{1,2}}=m^2_{D_{1,2}}=m^2_{E_{1,2}} 
=m_{3/2}^{2}(1- \frac{3}{2}(X^{2}_{0}+X^{2}_{3})).
\end{eqnarray}
The Yukawa coupling of the third family is 
present in the superpotential at trilinear order and hence there is a 
prediction for the associated 33 entry of the trilinear coupling 
$\tilde{A}$:
\begin{equation}
\label{trilinear33}
\tilde{A}_{33}=-\sqrt{3}m_{3/2}X_{1}e^{-i\alpha _{1}}Y_{33}
\end{equation}
In the absence of a 
complete theory for the smaller Yukawa couplings it is not possible to 
predict the structure of the associated soft trilinear couplings.  
However, we can
speculate on their form assuming that the smaller Yukawas have been
generated at the unification scale and can be treated as effective  
trilinear superpotential couplings (as in e.g. 
\cite{khalil}).\footnote{See \cite{lebedev2} for a more complete approach 
in which the small Yukawas as well as the soft trilinears are determined 
within perturbative heterotic string models.} In this case, the soft 
trilinear couplings take the form
\begin{equation}
\label{trilinear}
\tilde{A}_{ij}=
\left[
\begin{array}{ccc}
d_4Y_{11} & d_4Y_{12} & d_2Y_{13}\\
d_4Y_{21} & d_4Y_{22} & d_2Y_{23}\\
d_3Y_{31} & d_3Y_{32} & d_1Y_{33}
\end{array}
\right],
\end{equation}
with 
\begin{eqnarray}
d_1 & = & -\sqrt{3}m_{3/2}X_{1}e^{-i\alpha _{1}}=-M_{2}\\
d_2 & = & \sqrt{3}m_{3/2}(\frac{1}{2}X_{0}e^{-i\alpha _{0}}-
\frac{1}{2}X_{3}e^{-i\alpha _{3}})+d_1\\
d_3 & = & \sqrt{3}m_{3/2}(\frac{1}{2}X_{0}e^{-i\alpha _{0}}-
X_{2}e^{-i\alpha _{2}}+\frac{1}{2}X_{3}e^{-i\alpha _{3}})+d_1\\
d_4 & = & \sqrt{3}m_{3/2}(X_{0}e^{-i\alpha
_{0}}-X_{2}e^{-i\alpha_{2}})+d_1.
\end{eqnarray}
From Eq.~(\ref{trilinear}), the trilinear soft terms have
a form similar to Eq.~(\ref {yukorg}). 
The pattern of soft trilinear
couplings described above is similar to perturbative heterotic orbifold 
string-motivated models with matter in the twisted sector 
\cite{khalilref}.

The 
supersymmetric Higgs mass 
parameter $\mu$ is absent in the string-derived 
superpotential. In principle, $\mu$ and and the associated soft parameter 
$b$ must be generated in the 
effective theory (e.g. by the Giudice-Masiero mechanism \cite{gm}). 
As this is model-dependent, we prefer 
to leave them as free parameters in the phenomenological analysis (their 
magnitudes will be fixed by the requirement of radiative electroweak 
symmetry breaking).

\section{Phenomenological Implications}

\subsection{Approximate Unification of Gauge Couplings} 

The gauge couplings of string models are related to the size and
shape of the compactification manifold. Thus, the experimental
measurements of the gauge couplings provide information about the nature
of the compactification of a specific model. It is important to check that
the phenomenological requirements on the gauge couplings do not introduce
any inconsistencies.  Specifically, the mass scales
associated with the Kaluza-Klein (KK) and winding states 
introduce
corrections to the running of the gauge couplings.  As the 4224 
symmetry group breaks at the traditional 
unification scale $M_U \sim
10^{16}$ GeV, we choose to identify the lowest scale as $M_U$
so that the usual RG running 
in the
$d=4$ effective field theory can be applied. 
The gauge couplings of the five-branes are given by
\be
g_{5_i}^2=\frac{4 \pi}{ReT_i}= \frac{4 \pi \lambda_I M_i^2}{2 M_I^2},
\label{gradius}
\ee
where $M_i\propto 1/R_i$ characterizes the compactification size of the
$i$th compact complex dimension, $\lambda_I$ is the Type I string
coupling, and $M_I$ is the Type I string
scale. Defining $g_{5_2}=ag_{5_1}$, we obtain a relation 
between the compactification size of the two intersecting branes: 
\be
\frac{M_1}{M_2}=\frac{R_2}{R_1}=\frac{1}{a}. 
\ee
In the 4224 model, the gauge couplings of the Standard
Model gauge group near the unification scale are given by  Eq.~(\ref
{gaugecouplings1}):
\begin{eqnarray}
\label{gaugecoup}
g_3^2 &=&\frac{g_{5_1}^2 g_{5_2}^2}{g_{5_1}^2 + g_{5_2}^2}= 
\frac{g_2^2}{1 + 1/a^2}\nonumber \\
g_2^2 &=& g_{5_1}^2 \nonumber \\
g_Y^2 &=& \frac{3g_{5_1}^2 g_{5_2}^2}{3g_{5_1}^2 + 2g_{5_2}^2}=
\frac{3}{5}\frac{g_2^2}{1 + 2/5a^2}. 
\end{eqnarray}
The gauge couplings are only unified in the $a\rightarrow
\infty$ limit such that $T_2 \ll T_1$  ($R_2 \ll R_1)$.  This corresponds 
to a ``single brane'' limit in which the $5_2$ gauge sector decouples 
from the low energy physics, as can be seen from 
Eq.~(\ref{m3}).
We can investigate {\it approximate} gauge coupling unification by 
choosing $a^2$ rather large.
This effective ``single brane'' scenario will have important
phenomenological implications on the low energy spectrum of the model.

Notice that Eq.~(\ref{gradius}) is valid only at the string
scale, while the unification condition should be imposed at the
unification scale. For self-consistency, it is necessary that the string
and unification scales are very close to each other, such that the RG
running does not significantly modify
the ratio of the gauge couplings. An additional constraint on the model is  
the usual requirement that the theory remains in the
perturbative region ($\lambda_I < 1$). As an example,
we present a simple self-consistent realization. 
Consider the case in which all inverse radii except for
$R_2\equiv1/M_2$ are equal:
\be
M_1=\frac{M_2}{a}=M_3=M_c.
\label{gscenario}
\ee
Using the relations between the Planck scale, string scale, and 
compactification scales in Type I theory \cite{ibanez}, one can obtain 
\cite{rigolin} the useful 
relation $\alpha_{5_i} M_P/\sqrt{2} = M_I^2 M_i/(M_j M_k)$ $(i\neq
j \neq k)$.
Using this result and Eq.~(\ref{gscenario}),
the string scale and compactification
scale $M_c$ are related by 
\be
\frac{aM_I^2}{M_c} \simeq 3.5\times 10^{17} {\rm GeV}.
\label{scalerelation2}
\ee
Now consider the case in which $M_c< M_I/\sqrt{a}$. The lowest scale is a
winding mode with mass $M_c$ and thus we identify $M_c=M_U=3 \times 
10^{16}$ GeV. 
From Eq.~(\ref{scalerelation2}), 
the string scale (for $a=4$) is $M_I=5 \times 10^{16}$ GeV. The 
theory remains in
the perturbative regime, as $\lambda_I= 2 \sqrt{2} M_I^4/(a M_c^3 M_P)
<1$. Therefore, this approximate unification approach is consistent and
could in principle be realized in simple orbifold models.\footnote{A
related analysis with a similar conclusion was performed in
\cite{kingrayner}.}

Let us now investigate the implications of the approximate 
unification scenario (see \cite{leontaris} for a more detailed analysis 
for the nonsupersymmetric Pati-Salam brane models).  First, 
Eq.~(\ref{gaugecoup}) reduces $g_3$ compared to the naive unification, 
which is the direction required by experiment.\footnote{Note that in the 
embedding of the SM gauge group expected within the Shiu-Tye model (with 
the $SU(2)_{L,R}$ split between the two sets of branes), a similar 
analysis of the gauge  couplings leads to a prediction of $g_3$ in the 
opposite direction.} More precisely,
\begin{eqnarray}
\label{pred}
\sin^2(\theta_W)(M_Z)&=&\frac{3}{8}\frac{1}{1+1/4a^2}\left 
(1-\frac{\alpha (M_Z)}{2\pi}(b_Y+b_2)\log{\frac{M_U}{M_Z}}\right) 
+\frac{\alpha (M_Z)}{2\pi}b_2 \log{\frac{M_U}{M_Z}}\\
\frac{1}{\alpha_s(M_Z)}&=&\frac{3}{8}\frac{1+1/a^2}{1+1/4a^2} 
\left (\frac{1}{\alpha 
(M_Z)}-\frac{1}{2\pi}(b_Y+b_2)\log{\frac{M_U}{M_Z}}\right)
+\frac{1}{2\pi}b_3 \log{\frac{M_U}{M_Z}},
\end{eqnarray}
in which the MSSM beta functions are $b_Y=11$, $b_2=1$, $b_3=-3$. The 
standard unification picture is 
reproduced for larger values of $a$, but $a$ is bounded by the requirement 
that the gauge coupling of the $5_2$ branes remains 
perturbative.\footnote{Choosing $a^2$ much larger than $O(10)$ 
leads to $\alpha_{5_2}=g_{5_2}^2/4\pi 
\simgt O(1)$ for reasonable ranges of the initial value of 
$\alpha_{5_1}=g_{5_1}^2/4\pi$.}  
Acceptable values of $a$ tend to yield $\alpha_s(M_Z)$ slightly smaller 
than the preferred range. For $a=4$, Eq.~(\ref{pred}) 
yields $\sin^2(\theta_W)=0.23$ and $\alpha_s(M_Z)=0.11$.

\subsection{Electroweak Symmetry Breaking}

We now turn to an investigation of radiative electroweak symmetry breaking
within the 4224 model. As this model is a generalized Pati-Salam
model, the embedding of the matter content on the branes does not
distinguish between quarks and leptons (or between up and down type
quarks) of a single generation, as they all reside within a given
multiplet. The brane embedding is also such that third generation Yukawa
unification is obtained at the string scale (with Yukawa couplings of
$O(1)$), while the first and second generations get masses via
higher-dimensional operators.  Therefore, large $\tan \beta \equiv \langle
H_u \rangle / \langle H_d \rangle \simeq m_t / m_b \sim O(35-50)$ is
required to produce the correct spectrum of SM fermions.

Radiative electroweak symmetry breaking in models with large $\tan\beta$ 
has been discussed extensively in the literature 
\cite{yukawaunif}.
One can always choose within the bottom-up 
approach to fix $\mu$ and $b$ (or equivalently $B_{\mu}\equiv b/\mu$) by 
the 
electroweak symmetry breaking conditions. In the large $\tan \beta$ limit, 
the tree level minimization of the Higgs potential yields
\begin{eqnarray}
b \simeq \frac{1}{\tan \beta} [(m_{H_d}^2-m_{H_u}^2)-M_Z^2] \\
|\mu|^2 \simeq -m_{H_u}^2 - \frac{M_Z^2}{2}. 
\end{eqnarray}
The pseudoscalar mass-squared is given by
\be
m_A^2=\frac{b}{\sin 2\beta} \simeq
\frac{1}{2}[(m_{H_d}^2-m_{H_u}^2)-M_Z^2]. 
\ee
If the top and bottom Yukawas are comparable, then $m_{H_u}^2$ and
$m_{H_d}^2$ will run similarly, so both  $m_{H_u}^2$ and $m_{H_d}^2$ will
run to negative values.  Requiring the pseudoscalar mass-squared to be 
positive, we immediately obtain the result that 
$m_{H_d}^2=m_{H_u}^2=-M_Z^2$ is not allowed. 

Although this seems problematic because the input values of $m_{H_d}^2$
and $m_{H_u}^2$ are equal within models in which the electroweak
Higgs pair resides in a single GUT multiplet, these
input values can be split by D term contributions (for a general
discussion see \cite{koldamartin}) 
$m^2_{H_d}(M_U)-m^2_{H_d}(M_U)=4g_2^2D$.  These D term contributions to 
the soft masses of the electroweak Higgs doublets arise after the breaking 
of the Pati-Salam group and are proportional to the difference of the 
VEV's of 
the fields $H,\bar{H}$ which break the symmetry. As shown for 
Pati-Salam models \cite{MO}, the D terms can be nonzero and of order the 
soft breaking scale. 
This analysis utilized a particular superpotential
\cite{kingshafi}
which is symmetric under the 
exchange of $H$ and $\bar{H}$.
This symmetry between $H$ and $\bar{H}$ is broken by their soft masses, 
leading to a D term which is proportional to their difference.
In the 4224 model, since $H 
\rightarrow C^{5_1}_1$,
$\bar{H}\rightarrow C^{5_1}_2$, and $\varphi_{1,2}\rightarrow C^{5_15_2}$,
$D\simeq m_{3/2}^2(X_0^2-X_3^2)$.  (The 
$\varphi_{1,2}$ fields do not contribute to the D term because their soft 
masses are identical). 
Thus, $m^2_{H_d}(M_Z)- m^2_{H_u}(M_Z)\sim m_{3/2}^2 \gg M_Z^2$, 
consistent with radiative electroweak symmetry breaking.

While large $\tan\beta$ has certain desirable properties (such 
as large radiative corrections to the mass of the lightest neutral Higgs 
boson easily lift the mass to much larger values than the MSSM tree level 
result $m_h^0\simeq M_Z$), 
proper radiative electroweak symmetry breaking 
requires $\mu /\tan \beta \simeq B_{\mu}$, an apparent fine-tuning 
\cite{nelsonrandall}. 
Within the top-down 
approach,
this is a significant constraint in 
model-building.  For example, this relation seems unnatural within 
string-motivated supergravity models, for which $\mu\sim B_{\mu} \sim 
m_{3/2}$ unless some cancellation or special alignment of Goldstino 
angles is present.

\subsection{Low Energy Spectrum} 

In this model, the low energy spectrum of the MSSM states is governed by 
an important sum rule related to 
the choice of a high string scale $\simeq 10^{16}$ GeV.  In 
the gauge unification limit, $m_{3/2}^2 \simeq 
(m_3/\sqrt{3}X_1)^2 + O(1/a^2)$ (using Eq.~(\ref{m3})). From the 
expressions of soft masses and constraint
$\sum_{i=0}^{3} X_i^2=1$, the following sum rule at the
unification scale can be obtained: 
\be 
m_h^2+m_{\bar{F}_3}^2+m_{F_3}^2 = m_3^2 + O(1/a^2).  
\label{sumrule} 
\ee 
This sum rule is familiar from the form of the soft supersymmetry breaking
parameters within perturbative heterotic models.  From the expression for
the gaugino soft mass parameters, we see that by ``decoupling'' the
effects of the $5_2$ brane sector to achieve gauge coupling unification,
the gluino and bino masses approach the wino mass value
$\sqrt{3}m_{3/2}X_1e^{-i\alpha_1}$ that they would have if only the $5_1$
sector was present.  The soft parameters are similar
to those from perturbative heterotic models, which resemble
``single brane'' Type I scenarios.

To study the implications of this sum rule, we must take into account the
important requirement 
that the mass-squared eigenvalues of the sfermions
must be positive to avoid charge or color breaking (CCB) minima.  Positive 
input values for the soft superysmmetry breaking
mass-squared parameters of the sfermions are also required, which implies
that the Goldstino angles $X_2, X_3 < 1/\sqrt{3}$ at the unification
scale.  The squark mass-squared eigenvalues always satisfy the positivity
requirement because their diagonal terms run to larger values at low
energies (driven strongly by $m_3$). In the
slepton sector, given that $m_{\bar{F}_3}^2=m_{\tilde{\tau}_R}$ and
$m_{F_3}^2=m_{\tilde{\tau}_L}$ do not substantially run from $M_U$ to
$M_Z$ and the existence of large off-diagonal mixing for the third
generation sfermions, $m_{\bar{F}_3}^2$ and $m_{F_3}^2$ must be quite
large to satisfy the conditions for the absence of CCB minima. It is
a generic prediction (from Eq.~(\ref{sumrule})) that $m_3(M_U)$ must be
($\geq 350$) GeV. This yields a larger gluino mass $m_3(M_Z) \sim 1$ TeV,
disfavored from fine-tuning arguments \cite{finetune}.

In the large $\tan \beta$ limit, 
the tree level minimization condition of Higgs potential yields 
$|\mu|^2 \simeq -m_{H_u}^2(M_Z)-M_Z^2/2$.
The renormalization group running of the soft 
mass parameters yields the relations $m_{H_u}^2(M_Z)\simeq m_h^2(M_U) - 4 
m_3^2 (M_U)$ and $|\mu|^2 \simeq
3m_3^2(M_U)$. Therefore, proper radiative electroweak symmetry breaking 
requires $|\mu|\geq \sqrt{3}m_3(M_U) >
600$ GeV. This relatively large value of $\mu$ implies 
that the lightest two neutralino mass eignestates are bino and wino, with 
masses $m_{\chi^0_1}\simeq m_3(M_U)/2 \geq 
170$ GeV and $m_{\chi^0_2}\simeq m_3(M_U) \sim 350$ GeV.  The lightest 
chargino has the same mass as the neutral wino.     

The squarks are generally rather heavy, with the lightest state around 
$600$ GeV. $\tilde{\tau}_1$ is generally quite light 
due to the strong LR mixing.  In a significant fraction of the parameter 
space, the lightest stau can be the lightest supersymmetric particle.  
It is desirable though that 
the lightest superpartner is neutral and constitutes the cold 
dark matter (CDM) of the 
universe, which yields an additional 
constraint $m_{\tilde{\tau}} \geq m_{\chi^0_1} \simeq m_3(M_U)/2$.
However, the sum rule Eq.~(\ref{sumrule})  implies $m_{\bar{F}_3}^2 +
m_{F_3}^2 \leq  m_3(M_U)^2 \simeq 4m_{\chi^0_1}^2$. Taking into
account that the lightest mass of $\tilde{\tau}$ is a result of large
off-diagonal mixing, it is difficult to make its mass larger than 
the mass of the lightest neutralino. This constraint forces the 
condition $m_h \sim 0$ (i.e. $X_0\rightarrow 1/\sqrt{3}$).  
A particular choice of high energy parameters 
$X_0=0.58$,$X_1=0.79$, $X_2=0$, $m_{3/2}=318$ GeV, 
$\alpha_{1,2}=0$, $a=\sqrt{10}$, and $\tan\beta=35$ yields the result
$m_{\tilde{\tau}_1}=174$ GeV and $m_{\chi^0_1} = 171$ GeV. 
It is also important that the 
relic density produced by the LSP is within the range 
$\Omega h^2 \leq 0.5$. Assuming the LSP mass is below the top threshold, 
we have  checked that the relic density is below the experimental 
bound. The mass difference between the LSP and $\tilde{\tau}$ is 
very small so that coannihilation may become important. This will make the 
LSP relic density predicted in this model even smaller.

We now comment on the electric dipole
moment (EDM) constraints. If phases are
large and superpartner masses are small, the EDM's require cancellations
\cite{minn1,ibrahimnath,bgk,ckp,minn2,newedm,lebedev} which are
increasingly stringent for large $\tan\beta$.
Although the superpartner masses are rather large in this model, they are 
not 
large enough to allow $O(1)$ phases without cancellations. 
Assuming that the soft trilinear couplings are given by
Eq.~(\ref{trilinear}) and that the phases take the values
$\alpha_0=\alpha_1=\alpha_2=0$ ($\alpha_3$ general), the gaugino masses
are all real while only the 13, 23, 31, and 32 elements of the soft
trilinear couplings are complex. The EDMs are sensitive to the 11 entry of
the LR blocks in the SCKM basis. With this choice of phases (and the phase
of $\mu$ zero), the 4224 model is almost EDM-free because the 11 entries
of the LR blocks are real. The rotation of the soft trilinear matrices to
the SCKM basis does not reintroduce large CP violating phases to the 11
entry due to the factorizable structure of the $\tilde{A}$ terms
\cite{khalil,kv}.\footnote{However, it was recently pointed out that this 
feature does not necessarily hold
in generic string models (in conjunction with obtaining a large CKM
phase) \cite{lebedev2}.}

To summarize, the phenomenological pattern of the low energy mass spectrum
is related via the sum rule to the following set of theoretical
assumptions: (i) the high string and GUT breaking scales and (ii) only the 
MSSM states are present below the unification scale.
As the Type I string scale is flexible, a natural question to ask is what 
results if the string and symmetry breaking scales are lowered e.g. to 
intermediate values (recall that semirealistic brane models have been 
constructed with nonsupersymmetric as well as supersymmetric sectors, for 
which the string scale naturally should be of order $10^{11}$ GeV
\cite{ibanezquevedo,abelibanez,aldazabal2}).  Without a string-derived 
model, this depends on the assumed details of the symmetry
breaking scale and the low energy particle content.  In the 4224 model, 
the string scale cannot
be lowered to intermediate scale values with only MSSM matter 
content.  The model predicts $g^2_3=
g_2^2/(1+1/a^2)<g_2$ at the string scale, a condition violated by the
RG running using MSSM beta functions and the low energy values of the 
gauge couplings.
With additional matter there is flexibility to 
lower the scale while simultaneously lowering $a$ to deviate from the sum 
rule; this may be an interesting avenue for future exploration.

\section{Conclusions}

In this paper, we investigated aspects of the phenomenology of a Type I
string-motivated model with two intersecting (orthogonal)  stacks of
five-branes with Pati-Salam gauge groups.  The Pati-Salam $SU(4)_c$ is
obtained from the diagonal subgroup of the $U(4)$ gauge groups from each
set of branes, while $SU(2)_{L,R}$ are given by the $U(2)$'s from one of
the brane sectors. The third SM family is given by open string states from
the $5_1$ sector which have an $O(1)$ Yukawa coupling to the
electroweak Higgs doublets,
while the first and second families are intersection states with zero 
Yukawa
couplings at leading order. This 4224 model is similar to the Shiu-Tye
orientifold, but differs in the details of the SM gauge group embedding. 
The model has interesting features as well as generic 
problems which are summarized below.

An analysis of the diagonal breaking indicated that ensuring D flatness
for the $U(N)$ gauge groups of the theory requires a doubling of the usual
Higgs sectors needed for $SU(N)$ groups and predicts massless exotics at 
the renormalizable level.  Within the 4224 model,
this result leads to the presence of exotic colored states which render
the proton unstable if they remain massless at higher orders.  Decoupling
the additional triplet pair in the Pati-Salam symmetry breaking also is
more subtle within $U(N)$ groups because the usual trilinear operators are
forbidden by $U(1)$ gauge invariance.  These issues signify generic
challenges for string models which utilize intersecting D-branes in this
manner.

Within the 4224 model, identifying the string scale to be near the usual
GUT scale places significant constraints on the parameters of the theory.  
Obtaining proper low energy gauge couplings requires that one of the
branes essentially decouples from the low energy physics (hence there
is a hierarchy between the moduli VEVs which give the gauge couplings). 
In this effective single brane scenario, the soft terms are
similar to those derived within perturbative heterotic models and obey a
well-known sum rule. The constraints of the sum rule lead to a heavy
superpartner spectrum and thus are not favored by fine-tuning arguments,
but consistent radiative electroweak symmetry breaking can be achieved
using the D term contributions to the soft scalar masses of the Higgs
fields.  Such constraints may be alleviated if the string scale is 
lowered, at the price of additional matter content.

\acknowledgments
We thank M. Cveti\v{c}, G. Shiu, K. Dienes, and D. Chung for helpful
discussions. We thank the Aspen Center for Physics for hospitality during
several stages of this work.

\def\B#1#2#3{\/ {\bf B#1} (#2) #3}
\def\NPB#1#2#3{Nucl.\ Phys.\ B {\bf #1} (#2) #3}
\def\PLB#1#2#3{Phys.\ Lett.\ B {\bf #1} (#2) #3}
\def\PRD#1#2#3{Phys.\ Rev.\ D {\bf #1} (#2) #3}
\def\PRL#1#2#3{Phys.\ Rev.\ Lett.\ {\bf #1} (#2) #3}
\def\PRT#1#2#3{Phys.\ Rep.\ {\bf #1} (#2) #3}
\def\MODA#1#2#3{Mod.\ Phys.\ Lett.\ A {\bf #1} (#2) #3}
\def\IJMP#1#2#3{Int.\ J.\ Mod.\ Phys.\ A {\bf #1} (#2) #3}
\def\nuvc#1#2#3{ Nuovo Cimento\ {\bf #1}A (#2) #3}
\def\RPP#1#2#3{Rept.\ Prog.\ Phys.\ {\bf #1} (#2) #3}
\def\etal{{\it et al.\/}}

\bibliographystyle{prsty}

\end{document}